\documentclass[12pt]{article}

\usepackage{hyperref}
\usepackage{amssymb}

\begin{document}

%%%%%%%%%%%%%%%%%%%%%%%%%%%%%%%%%%
\renewcommand{\abstractname}{\hfill}

%%%% *****************************************************************
%%%% *************    Text stat'i     ********************************
%%%% *****************************************************************
\newpage
\pagenumbering{arabic}

{\Large \bf Gravitational Radiation as the Bremsstrahlung

\vspace{4pt}
of Superheavy Particles in the Early Universe
}

\vspace{11pt}
{\bf
Andrey A. Grib${}^{1,2,*}$ and Yuri V. Pavlov${}^{3,4}$
}
\begin{abstract}
${}^{1}$ Theoretical Physics and Astronomy Department, The Herzen  University,
48~Moika, St.\,Petersburg, 191186, Russia

${}^{2}$ A. Friedmann Laboratory for Theoretical Physics, St.\,Petersburg, Russia

${}^{3}$ Institute of Problems in Mechanical Engineering of
Russian Academy of Sciences,
61 Bolshoy, V.O., St.\,Petersburg, 199178, Russia; yuri.pavlov@mail.ru

${}^{4}$ N.I.\,Lobachevsky Institute of Mathematics and Mechanics,
Kazan Federal University, Kazan, 420008, Russia

${}^{*}$ Correspondence: andrei\_grib@mail.ru
\end{abstract}

    \begin{abstract}
\noindent
{\bf Abstract:}
    The number of superheavy particles with the mass of the Grand Unification
scale with trans-Planckian energy created at the epoch of superheavy particle
creation from vacuum by the gravitation of the expanding Universe is calculated.
    In later collisions of these particles gravitational radiation is radiated
playing the role of bremsstrahlung for gravity.
    The effective background radiation of the Universe is evaluated.

\vspace{7pt}
\end{abstract}

%%%% *****************************************************************
\section{\normalsize Introduction}
\label{secIn}

\hspace{\parindent}
    In paper published by one of the present authors (A.A.G.) together with
S.G. Mamayev~\cite{GribNuclPhys69} finite results for particle creation in
the early expanding Friedmann Universe were obtained.
    The important result was calculation of the finite density and finite
total number of particles created in the Lagrange volume.
    Our results were obtained by using Fock quantization with vacuum defined
as the ground state of the instantaneous Hamiltonian.
    These results have simple intuitive physical interpretation due to
the work made on the virtual pair on the Compton length of the particle by
the tidal forces due to gravity of the expanding Universe.
    It occurred that the number of particles depends on the mass of
the particle and leads to observable numbers of visible and dark matter
particles if this mass identified with the mass of the dark matter particles
is equal to the number close to the Great Unification scale~\cite{GMM}.
    Possible explanation of the origin of ultra high energy cosmic rays
due to the decay of such particles was made
in~\cite{GribPavlov2002(IJMPD)}--\cite{GribPavlov2007AGN}.

    So the main idea is that superheavy particles were created by gravitation,
then some part of them decayed on visible particles at high energies but
after the energy became smaller the decays were frozen and survived superheavy
particles formed observed dark matter~\cite{GribPavlov2002(IJMPD)}.
    However mathematical calculation of particle creation makes possible
calculation of the distribution function depending on the momenta i.e density
not in coordinate but in momentum space.
    Surely finiteness of the particle density in coordinate space means
that this function is going to zero at very high values of the momentum.
    This means that the larger the momentum the smaller will be the number
of created particles.

    In this paper we shall answer the question about the number of superheavy
particles with Grand unification mass but with the energy close to
the Planckian mass.
    How many trans-Planckian particles are created?
    Surely their number is much smaller than the general number of created
particles but how much?
    For the case of inflationary models
similar question was asked for same particles called
wimpzillas~\cite{KolbStarob07}.
    Why this question and the answer on it are important?
    It is because in our recent paper~\cite{GribPavlov2020d} it was found that
at trans-Planckean energies of colliding particles the gravitation wave
radiation is produced.
    This radiation is much stronger than comparable electromagnetic radiation
if it could exist.
    It plays the role of bremsstrahlung for superheavy particle and can lead
to formation of structures for them as it is the case for electromagnetic
radiation at small energies when Galaxies, stars etc. can be formed.
    In paper~\cite{Hooft87} it was also shown that collision of particles with
trans-Planckean energies can lead to formation of mini black holes.
    The problem of the formation of primary galactic nuclei during phase
transition in the early Universe was considered in~\cite{RKHlopov01}.

    In the book~\cite{Grib95} it was mentioned that in Friedmann expanding
Universe the number of created particles is proportional to the number of
causally disconnected parts of the Universe at the Compton time of
the existence of the Universe.
    However later these disconnected parts are united and collisions of
particles and black holes occur.
    This makes possible the speculation of possible arising of the primordial
black holes leading to active nuclei of Galaxies by this mechanism.

\vspace{4mm}
%%%% *****************************************************************
\section{\normalsize Creation of Superheavy Particles with Planckian
Energy in the Early Universe}
\label{secRPlE}

\hspace{\parindent}
     Consider creation of scalar and spinor particles in the early
homogeneous isotropic Friedmann Universe with metric
    \begin{equation}
ds^2=g_{ik}dx^i dx^k = a^2(\eta)\,(d{\eta}^2 - d l^2) \,,
\label{gik}
\end{equation}
    where $d l^2 $ is the metric of an 3-dimensional space of constant
curvature $K=0, \pm 1 $.

    In theory of quantum effects in expanding curved space-time one usually
takes the following equation (in the system of units in which Planck constant
and light velocity are equal to one: $\hbar = c =1$) for scalar field
of mass~$m$
    \begin{equation}
( \nabla^i \nabla_{i} + \xi R + m^2 )\, \varphi(x)=0 ,
\label{Eqm}
\end{equation}
    corresponding to the Lagrangian
    \begin{equation}
L(x)=\sqrt{|g|} \left[\, g^{ik}\partial_i\varphi^*\partial_k\varphi -
(m^2 + \xi R)\, \varphi^* \varphi \right],
\label{Lag}
\end{equation}
    where $ g = {\rm det}(g_{ik})$ and $R$ is the curvature scalar~\cite{BD}.

    For $\xi=\xi_c \equiv 1/6 $ the scalar field is called conformal coupled.
    Then the equation~(\ref{Eqm}) is conformally invariant in massless
case~\cite{GMM}.
    However the nonconformal case is also important because
1) ``gravitons''~\cite{GrishchukY80},
2) longitudinal components of vector bosons~\cite{GMM} are nonconformal.
    Minimal coupling $\xi=0$ is popular in inflation theories~\cite{Linde}.

    We can find a complete set of solution for the equation~(\ref{Eqm})
in following form
    \begin{equation}
\varphi(x)  % = \frac{\tilde{\varphi}(x)}{a (\eta)}
= \frac{g_\lambda (\eta)}{a(\eta)} \Phi_J ({\bf x}),
\label{fgf}
\end{equation}
    where
    \begin{equation}
g_\lambda''(\eta)+\Omega^2(\eta)\,g_\lambda(\eta)=0 ,
\label{gdd}
\end{equation}
       \begin{equation}
\Omega^2(\eta)= \left( m^2 + (\xi - \xi_c) R \right) a^2 +\lambda^2 ,
\label{Ome}
\end{equation}
     \begin{equation}
\Delta \Phi_J ({\bf x}) = - \left( \lambda^2 - K \right) \Phi_J  ({\bf x}),
\label{DFlF}
\end{equation}
    the prime denotes a derivative with respect to conformal time $\eta$,
and indices $J$ is numbering the eigenfunctions of Laplace-Beltrami operator $\Delta$
in the space~${\bf x}$ with the metric~$d l^2$.

    According to the Hamiltonian diagonalization method~\cite{GMM}
the functions $g_\lambda(\eta)$ satisfy initial conditions:
    \begin{equation}
g_\lambda'(\eta_0)=i\, \Omega(\eta_0)\, g_\lambda(\eta_0) , \ \
\ |g_\lambda(\eta_0)|= \frac{1}{\sqrt{\Omega(\eta_0)}} .
\label{icg}
\end{equation}

%%%%%%%%%%%%%%%%%%%%%%%%%%%%%%%%%%%%%%%%%%
     The number of pairs of scalar particles created up to the moment $t$
in Lagrangian volume $a^3(t) $ of homogeneous isotropic Universe with
flat space sections ($K=0$) is equal~\cite{GMM}
    \begin{equation}
N(\eta) = \frac{1}{2 \pi^2 } \int \limits_0^\infty s_\lambda(\eta)\, \lambda^2 d \lambda,
\label{r1}
\end{equation}
    where
    \begin{equation}
s_\lambda(\eta) = \frac{\left| g'_\lambda (\eta ) -
i \Omega \, g_\lambda (\eta ) \right|^2}{4 \Omega } .
\label{Sgg}
\end{equation}
    Function $s_\lambda(\eta)$ defines the distribution in dimensionless
momentum~$\lambda$ particles created up to the moment~$\eta$.
    The ``physical'' momentum is~$ \lambda/a$.
    For ultrarelativistic particles $\lambda \gg ma$.
    To find the number of created ultrarelativistic particles one must find
the asymptotic of the function $s_\lambda(\eta)$ if $\lambda \to \infty$.

    Using~(\ref{gdd}), (\ref{icg}) one can see that function $s_\lambda(t)$
satisfies integral Volterra equation
    \begin{equation}
s_\lambda(\eta) = \frac{1}{2} \int_{\eta_0}^\eta d \eta_1\, w(\eta_1) \int_{\eta_0}^{\eta_1}
d \eta_2\, w(\eta_2) \left( 1 + 2 s_\lambda(\eta) \right) \cos \left[
2 \Theta(\eta_2, \eta_1) \right],
\label{ius}
\end{equation}
    where
    \begin{equation}
w (\eta) =\frac{\Omega'(\eta)}{\Omega}, \ \ \ \
\Theta(\eta_2, \eta_1) = \int_{\eta_1}^{\eta_2} \Omega(\eta) \, d \eta.
\label{wThet}
\end{equation}
    To find the asymptotic $s_\lambda(\eta)$ if $\lambda \to \infty $
one can confine  oneself to the first iteration of the integral
equation~(\ref{ius}) and to take into account that
$\Theta(\eta_2, \eta_1) \to \lambda (\eta_1 - \eta_2 )$
if $\lambda \to \infty$.
    Then~(\ref{ius}) has the form
    \begin{equation}
s_\lambda(\eta) = \frac{1}{4} \left|
\int_{\eta_0}^\eta w(\eta_1) \exp( 2 i \lambda \eta_1 )\, d \eta_1 \right|^2.
\label{iusA}
\end{equation}
    Integrating by parts the integral~(\ref{iusA}) one obtains
in $O(\lambda^{-2})$
    \begin{equation}
\int_{\eta_0}^\eta w(\eta_1) \exp( 2 i \lambda \eta_1 )\, d \eta_1 \approx
\left. \frac{w(\eta)}{2 i \lambda} e^{2 i \lambda \eta} \right|_{\eta_0}^\eta .
\label{iusAk}
\end{equation}
    So
    \begin{equation}
s_\lambda (\eta) \approx \frac{1}{16 \lambda^2} \left| w^2(\eta) + w^2(\eta_0)
- 2 w(\eta) w(\eta_0) \cos (2 \lambda (\eta - \eta_0 ) \right|.
\label{slb}
\end{equation}
    For $\lambda \to \infty$ one has that $w \sim \lambda^{-2} $, so
$s_\lambda \sim \lambda^{-6} $ and the integral in~(\ref{r1}) is convergent.
    So in the method of Hamiltonian diagonalization the number of created
by gravitation scalar particles is finite as for conformal as for nonconformal
coupling with curvature.

    Evaluate the number of created particles with the energy $ E \ge E_b$,
where $ E_b \gg mc^2 $.
    The cyclic frequency $\Omega$ for the scalar field with conformal coupling
with curvature is
    \begin{equation}
\Omega(\eta) = \omega(\eta) \equiv \sqrt{ m^2 a^2 (\eta) + \lambda^2 }.
\label{omeg}
\end{equation}
    In this case
    \begin{equation}
w (\eta) =\frac{a'/a}{1 + \left( \frac{\lambda}{ma}\right)^2} .
\label{wThetCF}
\end{equation}
    For large~$\lambda$
    \begin{equation}
\lambda \gg ma \ \ \Rightarrow \ \
w \approx \frac{m^2 a' a}{ \lambda^2} .
\label{lBmaN}
\end{equation}
    Let us consider the situation when $a' a $ is increasing in time
and $ a'(\eta_0) a(\eta_0) \ll a'(\eta) a(\eta) $.
    Then
    \begin{equation}
s_\lambda (\eta) \approx \frac{w^2(\eta)}{16 \lambda^2}
= \frac{m^4  a^2 a^{\prime\,2}}{16 \lambda^6}, \ \ \lambda \to \infty.
\label{slbN}
\end{equation}

    Find what limitations on background matter of the Universe arise due to
this condition.
    Einstein's equations
    \begin{equation}
R_{ik} - \frac{1}{2} g_{ik} R = - 8 \pi G\, T_{ik} ,
\label{GR70Ein}
\end{equation}
    for homogeneous isotropic space in metric~(\ref{gik}) are
    \begin{equation}
\frac{1}{a^2} \left( \left(\frac{a'}{a}\right)^2 +K \right)
= \frac{8 \pi G\, \varepsilon }{3} \,,
\label{GR70Ein1}
\end{equation}
    \begin{equation}
- \frac{1}{a^2} \left[\ \left(\frac{a'}{a}\right)^{\, \prime} +
\frac{1}{2} \left( \left(\frac{a'}{a}\right)^2 +K \right)
\right] = 4 \pi G p ,
\label{GR70Ein2}
\end{equation}
    where $\varepsilon$ and $p$ are the energy density and pressure for
background matter.
    From~(\ref{GR70Ein1}), (\ref{GR70Ein2}) one has
    \begin{equation}
(a' a)' = \frac{a^2}{2} \left[ 3 \left( 1- \frac{p}{\varepsilon} \right)
\left(\frac{a'}{a}\right)^2 - \left( 1+ 3 \frac{p}{\varepsilon} \right) K \right].
\label{Einw}
\end{equation}
     For realistic models of the Universe in the epoch when the effects of
particle creation are important (the Compton time of the particle~\cite{GMM})
one can neglect the space curvature and consider $|a'/a| \gg 1$.
    So
    \begin{equation}
(a' a)' \approx \frac{3}{2} \left( 1- \frac{p}{\varepsilon} \right)
a^{\prime\,2}
\label{EinwN}
\end{equation}
    and $a' a$ is increasing if $p< \varepsilon$.
    Important case of dust $p=0$, $a= a_1 \eta = a_0 t^{2/3}$
and radiation dominated Universes
$p=\varepsilon/3$, $a= a_1 \eta = a_0 \sqrt{t}$
are in this family.
    Here $t$ is the coordinate time $dt = a d \eta$.
    The limiting case with scale factor $a= a_1 \sqrt{\eta} = a_0 t^{1/3}$
corresponds to the most rigid state equation $p = \varepsilon$.

    Using~(\ref{slbN}) one obtains that the number of particle pairs with
momenta larger then some $\lambda_b$, created in volume $a^3(t) $ of
homogeneous isotropic Universe is equal to
    \begin{equation}
N_b(\eta) \approx \frac{1}{32 \pi^2 } \int \limits_{\lambda_b}^\infty
\frac{m^4 a^2(\eta) a^{\prime\,2}(\eta)\, d \lambda}{ \lambda^4 }
= \frac{m^4 a^2(\eta) a^{\prime\,2}(\eta)}{96 \pi^2 \lambda_b^3 }.
\label{r1b}
\end{equation}
%%%%%%%%%%%%%%%%%%%%%%%%%%%%%%%%%%%%%%%%%%%%%%%%%
    Going to usual unit system is made by
    \begin{equation}
m \to \frac{mc}{\hbar} , \ \ \ \ a' \to \frac{\dot{a} a}{c},
\label{zam}
\end{equation}
    where the dot above symbols is the derivative with respect to
time~$t$.
    For ultrarelativistic particles with $ E_b= \lambda_b \hbar c / a $
and so~(\ref{r1b}) in usual units is
    \begin{equation}
N_b(t) \approx \frac{m^4 c^5 a \dot{a}^2}{96 \pi^2 \hbar\, E_b^3 }.
\label{r1bN}
\end{equation}
    For the scale factor $a \sim t^\alpha$ one obtains
    \begin{equation}
N_b(t) \approx
\frac{\alpha^2 m^4 c^5 a^3 }{96 \pi^2 \hbar\, E_b^3 t^2}.
\label{r1b2}
\end{equation}

    For Planckian energy $ E_b = E_{Pl} \equiv \sqrt{\hbar c^5 /G}$,
where $G$ is the gravitational constant,
and the Compton time $t = t_C \equiv \hbar/ (mc^2)$ one obtains
    \begin{equation}
E_b = E_{Pl} \ \ \Rightarrow \ \
N_b(t_C) \approx \frac{\alpha^2}{96 \pi^2}\left( \frac{l_{Pl}}{l_{C}} \right)^3
\left( \frac{a(t_C)}{l_{C}} \right)^3,
\label{r1b3}
\end{equation}
    where $l_C= \hbar/ (mc)$ is the Compton length of the particle,
    $l_{Pl}= \sqrt{\hbar G/c^3} $ is the Planckian length.
    For radiation dominated case ($\alpha=1/2$) one obtains
    \begin{equation}
N_b(t_C) \approx \frac{1}{384 \pi^2}\left( \frac{l_{Pl}}{l_{C}} \right)^{3/2}
\left( \frac{a(t_{Pl})}{l_{C}} \right)^3.
\label{r1bf}
\end{equation}

    Note that this result is valid not only for conformal but also for
nonconformal scalar particles because for the radiation dominated case~$R=0$
and formula~(\ref{omeg}) for the frequency~(\ref{Ome}) is correct for
any value~$\xi$ of the parameter of connection of the scalar field
with curvature.

    For observable Friedmann radiation dominated Universe one obtains
$a(t_{Pl}) \approx 10^{-5}$\,m
so that for the scale of Grand Unification $ m =10^{15} $\,GeV
one has $10^{67}$ particle with the energy of the Planckian order.
    The general number of all superheavy particles created in the early
Universe is close to the Eddigton number $10^{80}$~\cite{GMM}.
    So the number of trans-Planckian scalar particles is relatively small.

    Now consider the case of creation fermion particles.
    In this case the number of created pairs is~\cite{GMM}
    \begin{equation}
N^{(1/2)}(\eta) = \frac{1}{\pi^2 } \int \limits_0^\infty s^{(1/2)}_\lambda(\eta)\,
\lambda^2 d \lambda,
\label{s1}
\end{equation}
    but the expression for $s^{(1/2)}_\lambda$ does not coincide with the
formula~(\ref{Sgg}).
    However the first iteration in integral equation for
function~$s_\lambda(\eta)$ coincides with formular~(\ref{iusA})
if $w$ for the spinor field is
    \begin{equation}
w^{(1/2)} = \frac{m a' \lambda}{\omega^2} .
\label{s2}
\end{equation}
    For large values $\lambda$
    \begin{equation}
\lambda \gg ma \ \ \Rightarrow \ \ w^{(1/2)} \approx \frac{m a'}{ \lambda} ,
\ \ \ s^{(1/2)}_\lambda \approx \frac{m^2 a^{\prime\,2}}{ 16 \lambda^4}.
\label{s3}
\end{equation}
    The number of pairs of spinor particles with momenta larger
then some $\lambda_b$, created in volume $a^3(t) $ of homogeneous isotropic
universe is equal to
    \begin{equation}
N^{(1/2)}_b(\eta) \approx \frac{1}{16 \pi^2 } \int \limits_{\lambda_b}^\infty
\frac{m^2 a^{\prime\,2}(\eta)\, d \lambda}{ \lambda^2 }
= \frac{m^2 a^{\prime\,2}(\eta)}{16 \pi^2 \lambda_b },
\label{r1bAs}
\end{equation}
    in usual units
    \begin{equation}
N^{(1/2)}_b(t) \approx \frac{m^2 c a \dot{a}^2}{16 \pi^2 \hbar E_b }.
\label{r1bfuu}
\end{equation}
    For the scale factor $a \sim t^\alpha$ one obtains that the number of
the fermion created pairs is
    \begin{equation}
N^{(1/2)}_b(t) \approx \frac{\alpha^2 m^2 c a^3}{16 \pi^2 \hbar E_b t^2}.
\label{r1bfsf}
\end{equation}
    For Planckian energy $ E_b = E_{Pl} $ and the Compton time one obtains
    \begin{equation}
E_b = E_{Pl} \ \ \Rightarrow \ \
N^{(1/2)}_b(t_C) \approx \frac{\alpha^2}{16 \pi^2} \frac{ l_{Pl} a^3(t_C)}{l_{C}^4},
\label{r1b3f}
\end{equation}
    in particular, for radiation dominated case ($\alpha=1/2$) one obtains
    \begin{equation}
N^{(1/2)}_b(t_C) \approx \frac{1}{64 \pi^2}
\frac{ a^3(t_{Pl})}{l_{Pl}^{1/2}\, l_{C}^{5/2}}.
\label{r1bff}
\end{equation}
    For observable Friedmann radiation dominated Universe one obtains
$a(t_{Pl}) \approx 10^{-5}$\,m
so that for the scale of Grand Unification $ m =10^{15} $\,GeV
one has $10^{75}$ fermion particle with the energy of the Planckian order.

\vspace{4mm}
%%%% *****************************************************************
\section{\normalsize Gravitational Radiation in Collisions of Trans-Planckian
Superheavy Particles}
\label{sec3}

\hspace{\parindent}
    In our paper~\cite{GribPavlov2020d} there was obtained the following
result for the gravitational radiation energy in two-particle collision
with the energy $E_{\rm c.m.} < E_{Pl}$ in the center of mass system
    \begin{equation}
E = \frac{4}{\pi} \frac{E_{\rm c.m.}^{\,3}}{E_{Pl}^{\,2}}
\ln \left( \frac{E_{\rm c.m.}}{ E_{Pl}}
\frac{ M_{Pl}}{\sqrt{m_1 m_2}} \right),
 \label{GIWoI}
\end{equation}
    where $ M_{Pl} = \sqrt{\hbar c/G} =2.18\cdot 10^{-8}$\,kg is Planck mass,
$m_1$ and $m_2$ are the masses of colliding particles.
    Note that even for such light particles as electrons one
has $\ln (M_{Pl}/ m) < 52 $.
    So one can see from~(\ref{GIWoI}) that if $ E_{\rm c.m.}  \ll E_{Pl} $,
then gravitational radiation is small
    \begin{equation} \label{GIWoIo}
E_{\rm c.m.} \ll E_{Pl} \ \ \Rightarrow \\
\frac{E}{E_{\rm c.m.}} \ll 1 .
\end{equation}
    However for $ E_{\rm c.m.}  \ge E_{Pl} $ the result is different.
    One has
    \begin{equation} \label{pnn}
E \approx \frac{4}{\pi} E_{\rm c.m.} \ln \left( \frac{E_{\rm c.m.}}{ E_{Pl}}
\frac{ M_{Pl}}{\sqrt{m_1 m_2}} \right).
\end{equation}
    This means that the role of gravitational radiation becomes large
and it can play the role of bremsstrahlung in electromagnetic radiation.
    In~\cite{GribPavlov2020d} it was shown that if colliding particles have
nonzero electric charges then bremsstrahlung due to electromagnetism is much
smaller then the gravitational one.
    Let us evaluate the order of the energy of this gravitational radiation

    Evaluate the gravitational background radiation of these particles
supposing that the energy of gravitation is radiated in collisions at
Compton epoch $t \approx \hbar /m c^2 $ of the expanding Universe.
    Taking into account the decrease of the energy of gravitational waves
due to the expansion of the Universe one obtains in the modern
epoch $t_{\rm mod} \approx 10^{18}$\,s the value of the energy as
$10^{67} E_{Pl} \sqrt{t_C/t_{\rm mod}} \approx 10^{48} $\,J for scalar particles
and $\approx 10^{56} $\,J for fermion particles.
    So the density of this energy of radiation $\approx 10^{-30}$\,J/m$^3$
for case of scalar particles or $\approx 10^{-22}$\,J/m$^3$
for case of fermion particles
is much less then the energy of electromagnetic background radiation
equal to $\approx 4 \cdot 10^{-14}$\,J/m$^3$.

\vspace{7pt}
\noindent
{\bf Author Contributions:}
The authors have contributed equally to all parts of this work.

\vspace{7pt}
\noindent
{\bf Funding:}
This research is supported by the Russian Foundation for Basic research (Grant No. 18-02-00461 a).
The work of Yu.V.P. was supported by the Russian Government Program of Competitive Growth
of Kazan Federal University.

\vspace{7pt}
\noindent
{\bf Conflicts of Interest:}
The authors declare no conflict of interest.

%%%% *****************************************************************


\begin{thebibliography}{99}

\bibitem{GribNuclPhys69}
Grib, A.A.; Mamayev, S.G.
{On field theory in Friedmann space}.
{\em Yadernaya Fizika} {\bf 1969}, {\em 10}, 1276--1281 (1969)
[Engl. transl. in
{\em Soviet Journal of Nuclear Physics} {\bf 1970}, {\em 10}, 722--725].

\bibitem{GMM}
Grib, A.A.; Mamayev, S.G.; Mostepanenko, V.M.
{\em Vacuum Quantum Effects in Strong Fields};
Friedmann Lab. Publ.: St.Petersburg, 1994.

\bibitem{GribPavlov2002(IJMPD)}
Grib, A.A.; Pavlov, Yu.V.
Superheavy particles in Friedmann cosmology and the dark matter problem.
\href{https://doi.org/10.1142/S0218271802001706}
{{\em Int. J. Mod. Phys. D} {\bf 2002}, {\em 11}, 433--436}.

\bibitem{KhlopovD03}
Dubrovich, V.K.; Khlopov, M.Yu.
Primordial pairing and binding of superheavy charged particles in the early
Universe.
\href{https://doi.org/10.1134/1.1581955}
{{\em JETP Lett.} {\bf 2003}, {\em 77}, 335--338}.
%% Pis'ma v ZhETF (2003) {\bf 77}, No.7, 403--406. astro-ph/0206138

\bibitem{KhlopovD04}
Dubrovich, V.K.; Fargion, D.; Khlopov, M.Yu.
Primordial bound systems of superheavy particles as the source of ultra-high
energy cosmic rays.
\href{https://doi.org/10.1016/j.astropartphys.2004.07.002}
{{\em Astropart. Phys.} {\bf 2004}},
\href{https://doi.org/10.1016/j.astropartphys.2004.07.002}
{{\em 22}, 183--197}.

\bibitem{GribPavlov2007AGN}
Grib, A.A.; Pavlov, Yu.V.
Do active galactic nuclei convert dark matter into visible particles?
\href{http://dx.doi.org/10.1142/S0217732308027072}
{{\em Mod. Phys. Lett. A} {\bf 2008}, {\em 23}, 1151--1159}.

\bibitem{KolbStarob07}
Kolb, E.W.; Starobinsky, A.A.; Tkachev, I.I.
Trans-Planckian wimpzillas.
\href{https://doi.org/10.1088/1475-7516/2007/07/005}
{{\em JCAP} {\bf 2007}},
\href{https://doi.org/10.1088/1475-7516/2007/07/005}
{{\em 07}, 005}.

\bibitem{GribPavlov2020d}
Grib, A.A.; Pavlov, Yu.V.
On the limiting energy of the collision of elementary particles close to
horizon of the rotating black hole.
\href{https://doi.org/10.1142/S0217732320502624}
{{\em Mod. Phys. Lett. A} {\bf 2020}, {\em 35}, 2050262}.

\bibitem{Hooft87}
Hooft, G.'t.
{Graviton dominance in ultra-high-energy scattering}.
\href{https://doi.org/10.1016/0370-2693(87)90159-6}
{{\em Phys. Lett.~B} {\bf 1987}},
\href{https://doi.org/10.1016/0370-2693(87)90159-6}
{{\em 198}, 61--63}.

\bibitem{RKHlopov01}
Rubin, S.G.; Sakharov, A.S.; Khlopov, M.Yu.
The formation of primary galactic nuclei during phase transitions in
the early Universe.
\href{https://doi.org/10.1134/1.1385631}
{{\em  J. Exp. Theor. Phys.} {\bf 2001}, {\bf 92}, 921--}929.

\bibitem{Grib95}
Grib, A.A. {\em Early Expanding Universe and Elementary Particles};
Friedmann Lab. Publ.: St. Petersburg, 1995.

\bibitem{BD}
Birrell, N.D.; Davies, P.C.W.
{\em Quantum fields in curved space};
Cambridge University Press, Cambridge, 1982.

\bibitem{GrishchukY80}
Grishchuk, L.P.;  Yudin, V.M.
{Conformal coupling of gravitational wave field to curvature}.
\href{http://dx.doi.org/10.1063/1.524541}
{{\em J. Math. Phys.} {\bf 1980}, {\em 21}, 1168--1175}.

\bibitem{Linde}
Linde, A.D.
{\em Particle Physics and Inflationary Cosmology};
Harwood Academic, New York, 1990.

\end{thebibliography}
\end{document}